\documentclass{Interspeech}

\interspeechcameraready

\title{Generalizable Audio Spoofing Detection using Non-Semantic Representations}

\author[affiliation={1,3}]{Arnab}{Das}
\author[affiliation={1,3}]{Yassine}{El Kheir}
\author[affiliation={1}]{Carlos}{Franzreb}
\author[affiliation={1,2}]{Tim}{Herzig}
\author[affiliation={1,3}]{Tim}{Polzehl}
\author[affiliation={1,2}]{Sebastian}{Möller}

\affiliation{}{German Research Center for Artificial Intelligence}{Germany}
\affiliation{}{Technical University Berlin}{Germany}
\affiliation{}{Gretchen AI GmbH}{Germany}
\email{\{arnab.das, yassine.el\_kheir, carlos.franzreb,tim.herzig,tim.polzehl\}@dfki.de,sebastian.moeller@tu-berlin.de}

\usepackage{comment}
\usepackage{multirow}
\usepackage[subtle]{savetrees}
\begin{document}

\maketitle

\begin{abstract}
Rapid advancements in generative modeling have made synthetic audio generation easy, making speech-based services vulnerable to spoofing attacks. 
Consequently, there is a dire need for robust countermeasures more than ever.
Existing solutions for deepfake detection are often criticized for lacking generalizability and fail drastically when applied to real-world data.
This study proposes a novel method for generalizable spoofing detection leveraging non-semantic universal audio representations.
Extensive experiments have been performed to find suitable non-semantic features using TRILL and TRILLsson models.
The results indicate that the proposed method achieves comparable performance on the in-domain test set while significantly outperforming state-of-the-art approaches on out-of-domain test sets.
Notably, it demonstrates superior generalization on public-domain data, surpassing methods based on hand-crafted features, semantic embeddings, and end-to-end architectures.
\end{abstract}

\section{Introduction}
\label{sec:intro}
In the past decade, significant strides in generative speech research, particularly in the context of text-to-speech (TTS) and voice conversion (VC) systems, have enabled the generation of synthetic speech that is more natural sounding and of higher quality.
These developments have effectively narrowed the distinction between authentic (real) and artificial (synthetic) speech.
This development makes speech-based systems like automatic speaker verification (ASV) extremely susceptible to spoofing or more generally to \textit{presentation} attacks \cite{wang2020asvspoof}.
Furthermore, new TTS and VC systems are emerging so quickly that countermeasure systems are quickly rendered ineffective against a newer spoofing system.
Hence, research into accurate, robust, and generalizable countermeasure systems has become imperative and is of considerable interest to both academic and industrial sectors.

To advance this critical research domain, several challenges have been organized, including the ASVspoof\footnote{https://www.asvspoof.org/} and the Audio Deep Synthesis Detection (ADD)\footnote{http://addchallenge.cn/} challenges, which have gained popularity.
These challenges provide essential datasets to develop countermeasure systems.  
The datasets include spoofed and synthetic speech generated by a wide range of TTS or VC systems as well as synthetic speech generated using heuristic or statistical methods. 
In contrast to the controlled datasets commonly associated with research challenges, which predominantly feature meticulously curated laboratory data, researchers have also proposed other datasets like ``In the Wild" (ItW) \cite{muller2022does}.
This dataset contains uncontrolled, noisy bonafide, and fake data directly from the public domain, which is not part of any such challenge and better reflects real-world scenarios.

\begin{figure*}[ht]
  \centering
  \includegraphics[width=\textwidth]{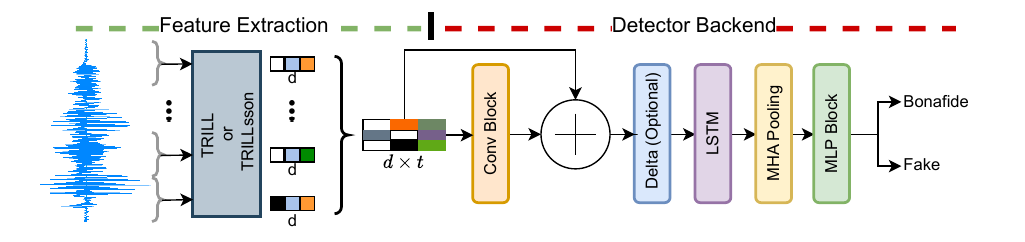}
  \caption{Schematic architecture diagram of the proposed framework.}
  \label{fig:speech_production}
\vspace{-0.5cm}
\end{figure*}

Literature on synthetic or spoofed speech detection formulated the task as a binary classification problem \cite{chen2017resnet, li2021replay, zhang2021fake} between the bonafide and fake classes.
Several attempts have been made over the years to improve the accuracy of this classification task by proposing different architectures and training strategies, namely end-to-end, hand-crafted feature-dependent, and self-supervised semantic feature-based models \cite{yi2023audio}.
In early countermeasure frameworks, classifiers such as the Gaussian mixture model (GMM) \cite{wang2015relative} or support vector machine (SVM) were trained on hand-crafted speech representations like linear frequency cepstral coefficient (LFCC) \cite{chen2021ur} and mel-frequency cepstral coefficient (MFCC) \cite{sahidullah15_interspeech}.
Early DNN-based methods by \cite{hua2021towards} introduced ResNet and Inception-based binary classifiers, outperforming prior speech representation frameworks. 
The authors suggest that shallow networks are better suited for detecting artifacts from spoofing algorithms, rather than focusing on high-level, respectively deep semantic speech structures.
In \cite{mo2022multi}, a multi-task learning objective is proposed and demonstrated to outperform traditional ResNet methods.
An alternative end-to-end framework leveraging RawNet2 \cite{tak2021end} directly processes raw waveforms, utilizing a fixed band-pass filter bank modeled as sinc functions \cite{ravanelli2018speaker}, followed by a trainable convolutional neural network (CNN).
Subsequently, an advanced variant of RawNet2 is proposed in \cite{li2023advanced}, which employs an attention-based channel masking mechanism and claims to improve performance. 

Recently, \cite{jung2022aasist} proposed AASIST, a graph neural network-based method, which shows SOTA performance on ASVspoof challenge datasets.
Apart from these methods, some of the latest models also use semantic speech representations extracted from the audio and perform the spoofing detection task on the embedding space. 
A fusion method proposed in \cite{yang2024robust} uses XLS-R\cite{babu2021xls}, WavLM\cite{chen2022wavlm}, and Hubert\cite{hsu2021hubert} embeddings as multiple views of the same input waveform, whereas \cite{kulkarni2024exploring} uses WavLM features with Ecapa-tdnn\cite{desplanques2020ecapa} backend.
However, some of these countermeasure techniques have been criticized for drastically failing to generalize well to previously unseen test datasets \cite{muller2022does, muller2024mlaad}.

In contrast to semantic audio embeddings, TRILL \cite{shor2020towards}, a self-supervised universal non-semantic audio representation extraction model trained with TRIpLet loss, excels in non-semantic tasks such as language identification, speech command recognition, and emotion detection by discarding semantic temporal features like content. 
Subsequently, TRILLsson \cite{shor2022trillsson}, a series of universal paralinguistic embedding models trained via knowledge distillation, was introduced. 
Similar to TRILL, TRILLsson models achieve improved performance on non-semantic tasks outlined in the NOSS benchmark \cite{shor2020towards}, while being significantly smaller in terms of the number of parameters than semantic embedding models like XLS-R.

To address the challenges of robustness and generalizability in developing countermeasures, this study proposes a novel approach using TRILL/TRILLsson features and investigates the effectiveness of non-semantic representations for spoof detection by conducting cross-dataset evaluations.
Experimental results demonstrate performance comparable to state-of-the-art (SOTA) models for in-domain evaluation.
Moreover, the proposed method leveraging these representations exhibits improved generalization to out-of-domain test datasets. 
Notably, it achieves superior performance on noisy real-world public-domain data, surpassing SOTA models.

\section{Methodology}
\label{sec:methodology}

A high-level architecture diagram of the proposed framework is illustrated in Figure \ref{fig:speech_production}. 
Initially, the input audio waveform is chunked into frames and each chunk is processed through TRILL or TRILLsson models to extract audio representations.
The resulting frame-wise representations are then stacked to form a 2D representation \( X \in \mathcal{R}^{d \times t} \), where \( d \) denotes the embedding dimension and \( t \) represents the number of frames. 
TRILL and TRILLsson models generate a global representation vector by temporal aggregation from entire waveforms of any length. 
Therefore, chunking is necessary to obtain localized non-semantic feature vectors for more effective analysis.
The optimal balance between localized and globalized features necessary for spoofing detection tasks remains an open question and is subject to our experimental evaluation. 
In the proposed framework, TRILL and TRILLsson models serve as fixed feature extractors with frozen weights and are not fine-tuned. 
The pre-trained model weights are publicly available\footnote{https://www.kaggle.com/models/google/nonsemantic-speech-benchmark/tensorFlow2/trill}\textsuperscript{,}\footnote{https://www.kaggle.com/models/google/trillsson/}. 
Non-semantic representations from TRILL and TRILLsson models exhibit greater temporal stability than semantic embeddings, as non-semantic attributes of speech evolve slowly and gradually compared to rapidly fluctuating lexical and phonetic features \cite{shor2020towards}.
Given that, this study experiments with TRILL and TRILLsson variants 1 to 4.
The different TRILLsson models represent distinct network architectures, as outlined in \cite{shor2022trillsson}.
The embedding dimension \( d \) is set to 512 for TRILL-based models and 1024 for TRILLsson-based models.
Additionally, we experimented with various chunking window sizes (50ms, 100ms, 200ms, and 300ms) to determine the optimal window for spoofing detection.

The stacked representation \(X\) is processed through a convolutional block with a residual connection.
This block comprises two layers of 1D convolutional filters, followed by batch normalization and SELU activation \cite{klambauer2017self}.
The convolutional block extracts high-level features, while the residual connection preserves relevant low-level information. 
Hence this convolution block effectively augments the incoming feature vector \(X\) $\rightarrow$ \(X'\).
The features then optionally undergo a frame-wise delta step to capture variations in non-semantic features across consecutive frames, performing $X''_t = X'_{t+1} - X'_t$.
Experiments are conducted both with and without this step. 
Subsequently, two LSTM layers model long-term temporal dependencies or inconsistencies.
Post LSTM, the features are projected to a 1536-dimensional space. 
A multi-head attention (MHA) pooling mechanism is then applied along the time dimension to focus on salient regions of the sequence. 
Finally, an MLP block generates logits for the bonafide and fake classes.
The framework is trained using cross-entropy loss with class weights of 0.1 and 0.9, following the recommendations of \cite{jung2022aasist}.
The code is available online \footnote{https://github.com/arnabdas8901/TRILLFake}.
\begin{table}[ht]
\centering
\caption{Evaluation results (EER\%) of different TRILL or TRILLsson representation-based models on the LA19 evaluation set. Results considering different feature extraction window sizes are also reported. For each model, the best configuration is in boldface.}
\label{tab:compare_trill}
\begin{tabular}{|l|l|c|c|c|c|c|}
\toprule
\textbf{Models} & \textbf{Type} & \textbf{50ms} & \textbf{100ms} & \textbf{200ms} & \textbf{300ms} \\
\midrule
{\multirow{2}{*}{$M_T$}} & Direct   & 9.29 & 9.49 & 8.48 & \textbf{7.45} \\
 & Delta   & 8.8 & 9.15 & 8.86 & 8.77 \\
\hline
\multirow{2}{*}{$M_{T1}$} & Direct   & 8.46 & 3 & \textbf{1.39} & 2.56 \\
 & Delta   & 6.12 & 4.24 & 3.34 &  5.8\\
 \hline
\multirow{2}{*}{$M_{T2}$} & Direct   & 2.78 & 1.67 & 1.56 & \textbf{1.43} \\
 & Delta   & 3.16 & 2.29 & 1.93 & 1.85 \\
\hline
\multirow{2}{*}{$M_{T3}$} & Direct   & 2.54 & 1.35 & \textbf{0.96} & 1.32 \\
 & Delta   & 2.87 & 2.04 & 1.4 & 1.45 \\
 \hline
\multirow{2}{*}{$M_{T4}$} & Direct   & 13.24 & 11.41 & 4.49  & \textbf{2.39} \\
 & Delta   & 17.72 & 14.11 & 6.33 & 2.67 \\
 \bottomrule
\end{tabular}
 \vspace{-0.6cm}
\end{table}

\begin{table*}[ht!]
\centering
\caption{Spoofing algorithm wise EER(\%) score comparison for LA19 evaluation set.}
\label{tab:classwise_tab}
\begin{tabular}{|l|c|c|c|c|c|c|c|c|c|c|c|c|c|}
\toprule
\textbf{Models} & \textbf{A07} & \textbf{A08} & \textbf{A09} & \textbf{A10} & \textbf{A11} & \textbf{A12} & \textbf{A13} & \textbf{A14} & \textbf{A15} & \textbf{A16} & \textbf{A17} & \textbf{A18} & \textbf{A19} \\
\midrule
\textbf{$M_{T1}$} (200 ms) & 0.08 & 1.34 & 0.12 & 0.49 & 0.51 & 0.08 & 0.06 & 0.26 & 0.24 & 0.51 & 1.08 & 2.97 & 4.29 \\ 
\hline
\textbf{$M_{T3}$} (200 ms) & 0.77 & 2.44 & 0.34 & 0.85 & 0.67 & 0.38 & 0.08 & 0.55 & 0.71 & 0.42 & 0.53 & 1.09 & 2.52 \\ 
\hline
\textbf{$M_{T3}$} (50 ms) & 0.12 & 5.45 & 0.24 & 0.43 & 0.39 & 0.16 & 0.12 & 1.02 & 0.61 & 0.71 & 5.02 & 2.2 & 6.08 \\ 
\bottomrule
\end{tabular}
\vspace{-0.4cm}
\end{table*}

\section{Experimental setup}
\label{sec:exp_setup}
\subsection{Datasets}
\label{sec:data}
We conducted extensive experiments using four distinct English datasets.

\noindent\textbf{ASVspoof 2019 Logical Access (LA19):} 
ASVspoof 2019 challenge made this dataset publicly available \cite{wang2020asvspoof}, comprising $12,483$ genuine and $108,978$ spoofed utterances sampled at \SI{16}{\kilo\hertz}. 
The dataset features male and female speakers sourced from the VCTK corpus of read speech \cite{yamagishi2019cstr}. 
The spoof utterances are generated using 19 different algorithms (A01 - A19), including 17 different TTS and VC techniques. 
In the evaluation subset, counterfeit utterances result from a distinct set of spoofing attack algorithms (A07-A19), while spoof samples in the training and development datasets originate from the same set of spoofing algorithms (A01-A06). 
The LA19 training subset is used to train all our models.

\noindent\textbf{ASVspoof 2021 LA (LA21):} The dataset for the ASVspoof challenge in 2021 \cite{liu2023asvspoof} extends its predecessor, aiming to narrow the disparity between controlled laboratory conditions and real-world scenarios. 
This is achieved by introducing channel artifacts, wherein both bonafide and spoof samples undergo transmission through various real telephony systems along with different speech codecs applied to them. 
Although the number of speakers and spoofing algorithms is unaltered from the previous LA19 version, the LA 21 set contains additional bonafide and spoof speech samples.

\noindent\textbf{ASVspoof 2021 DF (DF21):} The deepfake (DF) evaluation set from the ASVspoof 2021 challenge \cite{liu2023asvspoof} includes 14,869 bonafide samples and 519,059 spoofed samples, sourced from 50 female and 43 male speakers. 
Unlike the LA sets, the DF set features source utterances also from the 2018 and 2020 Voice Conversion Challenge (VCC) databases.
Spoofed utterances in this DF set are processed with lossy codecs meant for media storage. 
The DF set also includes previously unseen voice conversion (VC) algorithms that are not present in any of the LA sets. 

\noindent\textbf{In the Wild (ItW):} The dataset comprises approximately $17.2$ hours of fake and $20.7$ hours of authentic audio clips (podcasts, speeches, etc.) featuring English-speaking celebrities or politicians, totaling $31,779$ utterances with an average duration of $4.3$ seconds \cite{muller2022does}. 
All data within this collection are sourced from the public domain.
The details of the spoofing algorithms used in this dataset are not provided.
Additionally, the utterances include background noise, which more accurately reflects realistic scenarios.

In this study, the LA21 evaluation, DF21 evaluation, and ItW datasets are used solely to evaluate the generalization capabilities of the proposed framework. 
No training or fine-tuning is conducted on these datasets.

\subsection{Implementation details}
The model based on TRILL features is referred to as \( M_{T} \), while the models based on representations extracted using TRILLsson 1, 2, 3, and 4 are denoted as \( M_{T1} \), \( M_{T2} \), \( M_{T3} \), and \( M_{T4} \), respectively.
In this study, the maximum audio length is set to 6 seconds.
Longer audio samples are truncated, while shorter ones are padded.
Our models are trained for 50 epochs on an NVIDIA H100 GPU with a batch size of 64. 
Best models are chosen based on performance on the LA19 dev set.
Optimization is performed using the Adam optimizer with an exponential learning rate scheduler, where the initial learning rate is set to \(10^{-4}\) and decays at a rate of $5\%$ after each epoch.
For evaluation, We report the equal error rate (EER).

\section{Results \& discussion}
\label{sec:result_dis}

To determine the most suitable representation for spoofing detection among TRILL and TRILLsson models, we evaluate our models on the LA19 evaluation set.
The results are presented in Table \ref{tab:compare_trill}. 
For each model, two rows are presented based on whether the frame-wise delta is performed or not as mentioned in Section \ref{sec:methodology}. 
We define that in the \textit{Type} column.
\textit{Direct} refers to the model configuration without the optional delta step.
We also experimented with different chunking window sizes for feature extraction.
We can gain a few insights from Table \ref{tab:compare_trill}.
1) For each model configuration, the best result is achieved when the representations are extracted using a 200ms or 300ms long window. 
2) For models $M_T$, $M_{T2}$, and $M_{T4}$, the performance kept on increasing with larger window size and eventually getting the best scores for 300ms chunk size.
We empirically found that further increasing the chunking window size does not improve results. 
The best possible score for $M_T$ is $7.45\%$ EER whereas the same for the other two models are $1.43\%$ and $2.39\%$ respectively.
Whereas for $M_{T1}$ and $M_{T3}$, the best possible results are achieved with representations from 200ms chunks.
The models achieve an EER score of $1.39\%$ and $0.96\%$ respectively.
These results show that TRILLsson features are better suited for the task compared to TRILL features.
This performance difference aligns with the superior effectiveness of TRILLsson models over TRILL in non-semantic tasks, as TRILLsson outperforms TRILL across all tasks in the NOSS benchmark \cite{shor2022trillsson}.
3) For $M_T$ and $M_{T1}$ models the Delta variant works better than the Direct variant when the representations are extracted using smaller chunk size for example 50ms. 
However, this trend is not followed for other models and for most of the configurations, the Delta variants are underperformant compared to the Direct ones.

Based on the results in Table \ref{tab:compare_trill}, the top two models, \( M_{T1} \) and \( M_{T3} \), both trained with a 200ms chunking size, are selected for further experiments.

The LA19 evaluation set includes diverse spoofing mechanisms, leading to varying performance across detection algorithms. To analyze this, we compare the performance of the two selected models against each spoofing algorithm, with results presented in Table \ref{tab:classwise_tab}.
For algorithms \( A07 \) to \( A15 \), \( M_{T1} \) outperforms \( M_{T3} \), whereas for spoofing methods \( A16 \) to \( A19 \), \( M_{T3} \) demonstrates superior performance. Notably, \( A17 \), \( A18 \), and \( A19 \) are exclusively voice conversion (VC) frameworks, while methods \( A07 \) to \( A15 \) involve text-to-speech (TTS) systems or a combination of TTS and other techniques \cite{wang2020asvspoof}.  
The performance difference between these two models may arise from variations in the features extracted by TRILLsson1 and TRILLsson3.
TRILLsson1 has only 5 million parameters, whereas TRILLsson3 contains 21.5 million parameters \cite{shor2022trillsson}, potentially leading to differences (shallow vs deep) in the learned representations.

\begin{table}[ht]
\centering
\caption{Comparison of EER(\%) for LA19, LA21, DF21 evaluation datasets achieved by different models only trained on LA19 training set. Best scores are in bold and second best scores are underlined.}
\label{tab:compare_models}
\begin{tabular}{|l|c|c|c|}
\toprule
\textbf{Models} & \textbf{LA19} & \textbf{LA21} & \textbf{DF21} \\
\midrule
LFCC \cite{wang2021investigating} & 2.98 & 20.93 & 23.05 \\
\hline
RawNet2 \cite{tak2021end} & 1.14 & 9.5 & 22.38 \\
\hline
RawGAT-ST \cite{kulkarni2024exploring} & 1.22 & 10.23 & 37.15 \\
\hline
AASIST \cite{kulkarni2024exploring} & \textbf{0.83} & 11.46 & 21.06 \\
\hline
ARawNet2 \cite{li2023advanced}
& 4.61 & 8.36 & 19.03 \\
\hline
SE-Rawformer \cite{liu2023leveraging} & 1.15 & \textbf{4.31} & 20.26 \\
\midrule
$M_{T1}$ (200ms, Direct) & 1.39 & \underline{6.36} & \underline{17.17} \\
\hline
$M_{T3}$ (200ms, Direct) & \underline{0.96} & 7.28 & \textbf{13.27} \\
\bottomrule
\end{tabular}
\vspace{-0.2cm}
\end{table}

Table \ref{tab:classwise_tab} also presents spoofing method-wise results for an alternative configuration of the \( M_{T3} \) model, trained on representations extracted from 50ms chunks.
Compared to the 200ms model, the 50ms model exhibits significantly inferior performance for \( A08 \), \( A17 \), and \( A19 \). 
This discrepancy may arise because embeddings extracted from 50ms chunks still could only capture very local temporal features, whereas certain spoofing methods, particularly generative models, introduce patterns at a global level \cite{yan2024df40}.
Consequently, representations derived from longer chunk sizes are better suited for detecting global temporal inconsistencies.

To assess the out-of-domain generalization capability of our proposed framework, we evaluate our models on LA21 and DF21 evaluation subsets without further fine-tuning and the results are summarized in Table \ref{tab:compare_models}.
We also compare our models against current SOTA methods from the literature.
For our models, the best score is reported from three different seed runs.
The RawNet2\cite{tak2021end} model achieves an EER score of $1.14\%$ on the LA19 evaluation set, SE-Rawformer \cite{liu2023leveraging} also archives a similar score of $1.15\%$ and AASIST \cite{jung2022aasist} achieves an even lower score of $0.83\%$.
Our models also perform similarly to the SOTA models, $M_{T1}$ and $M_{T3}$ achieve an EER score of $1.39\%$ and $0.96\%$ respectively.
Upon evaluation on the LA21 evaluation set, performance drops sharply for all SOTA models.
AASIST could only achieve $11.46\%$ EER and the same for RawNet2 is $9.5\%$. 
The best performance is achieved by SE-Rawformer, with an error rate of $4.31\%$.
Our framework demonstrates superior generalization ability compared to end-to-end models such as AASIST and RawNet2, as well as hand-crafted LFCC-based models. Notably, \( M_{T1} \) achieves an EER of $6.36\%$, while \( M_{T3} \) attains $7.28\%$.
Upon testing on an even more challenging DF21 evaluation set, the performances of the SOTA models deteriorate drastically.
SE-Rawformer achieves an EER of $20.26\%$, almost a five-times performance drop from LA21, whereas AASIST could only achieve $21.06\%$.
The limited generalization of models such as AASIST and RawNet2 has also been documented in several studies \cite{yi2023audio, muller2022does, kulkarni2024exploring}.
Under the same testing condition, $M_{T3}$ archives the best EER score of $13.27\%$, almost a 37\% performance improvement over AASIST, followed by $M_{T1}$ $17.17\%$.
This shows that models learned using our proposed framework with non-semantic representations hold the key to improved generalization toward unseen out-of-domain data for spoofing detection tasks.

\begin{table}[ht]
\centering
\caption{Comparison of out-of-domain evaluation for models trained on LA19 training set and tested on ItW.}
\label{tab:compare_ItW}
\begin{tabular}{|l|c|}
\toprule
\textbf{Models} & \textbf{EER(\%)$\downarrow$}\\
\midrule
RawGAT-ST \cite{kulkarni2024exploring}& 52.54 \\
AASIST \cite{kulkarni2024exploring}& 43.01 \\
RawNet2 \cite{muller2022does} & 33.94 \\
WavLM+Ecapa \cite{kulkarni2024exploring} & 34.64 \\
XLS-R,WavLM,Hubert \& Fusion \cite{zhang2024audio}& \underline{24.27} \\
\midrule
$M_{T1}$ (200ms, Direct) & \textbf{20.08} \\
$M_{T3}$ (200ms, Direct) & 27.52 \\
\bottomrule
\end{tabular}
\vspace{-0.3cm}
\end{table}

To further assess the generalization capability of the proposed framework on real-world data, we evaluate our models on the ItW dataset along with the SOTA models, and the results are presented in Table \ref{tab:compare_ItW}.
Performance by the SOTA models like AASIST drops drastically and could only manage to keep the EER at $43.01\%$.  
The spoofing detection method that uses semantic features like WavLm and Ecapa only achieves $34.64\%$ EER and the multi-view method that uses a fusion of different semantic features like XLS-R, WavLM, and Hubert achieves an EER of $24.27\%$
Whereas, our model $M_{T1}$ achieves the best EER score of $20.08\%$.
The results further confirm that utilizing non-semantic audio representations significantly enhances generalization to out-of-domain data, including noisy real-world utterances, compared to end-to-end models and approaches leveraging semantic embeddings.
\begin{table}[ht]
\centering
\caption{Ablation study: semantic vs non-semantic out-of-domain generalization performance (EER\%) by using same detector backend. }
\label{tab:comp_w2v2}
\begin{tabular}{|l|c|c|c|}
\toprule
Models & LA19 & LA21 & DF21\\
\midrule
$M_{T1}$ (200ms,Direct) & 1.39 & \textbf{6.36} & 17.17\\
$M_{T3}$ (200ms, Direct) & \textbf{0.96} & 7.28 & \textbf{13.27}\\
\midrule
XLS-R + Our Backend & 1.59 & 28.78 & 24.49\\
\bottomrule
\end{tabular}
\vspace{-0.5cm}
\end{table}

To ascertain the advantage of non-semantic features over semantic features, we conduct an ablation study by replacing TRILLsson features with Wav2Vec2-XLS-R\footnote{https://huggingface.co/facebook/wav2vec2-xls-r-300m} \cite{babu2021xls} features while keeping the detector backend unchanged. 
The results of this comparison are presented in Table \ref{tab:comp_w2v2}.
The model learned using Wav2Vec2-XLS-R features achieves $1.59\%$ EER on the LA19 evaluation set, similar to our $M_{T1}$ model. 
However, when tested of the LA21 evaluation set as part of out-of-domain evaluation the model fails drastically and shows a very high EER rate of $28.78$.
This finding aligns with the results reported in \cite{wang2021investigating}, which also observed a significant decline in generalization performance on LA21 when using fixed Wav2Vec2-XLS-R features.

This advantage may stem from the fact that most semantic feature extraction models employ a 25ms frame size with a 20ms stride, limiting them to capturing only localized features specifically trained to perform semantic tasks like automatic speech recognition.
Additionally, to preserve semantic meaning, these representations encode variations driven by rapidly changing lexical and tonal information overlooking some of the global residual artifacts from the generative spoofing algorithms. 
As a result, spoofing detection models built on these semantic features may overfit irrelevant features, yielding strong in-domain performance but failing to generalize effectively.
In contrast, our experiments demonstrate that temporally aggregated non-semantic features extracted over a longer window (i.e., 200ms) exhibit strong generalization across datasets. 
Notably, \cite{chen2022speechformer} reports that the maximum duration of phonemes in standard English speech is approximately 200ms, while \cite{campbell1989syllable} empirically finds the average syllable duration to be also around 200ms.
This gives us insight that features extracted over a duration of average syllable length are more beneficial for spoofing detection tasks. 
Similarly, end-to-end models that directly learn from raw waveform can overfit to irrelevant variations in the highly time variant signal.

\section{Conclusion \& outlook}
\label{sec:conclusion}
In this paper, we propose a spoofing detection method based on non-semantic audio representation extracted from TRILL and various TRILLsson models.
We perform extensive experiments to find the most suitable TRILLsson model and optimal chunking duration to balance the local and global temporal features. 
Results show that the proposed method with the most suitable configurations shows equivalent performance to SOTA methods in the case of in-domain testing.  
However, the proposed method outperforms existing methods on out-of-domain datasets proving the efficacy of non-semantic features for better generalization over hand-crafted, end-to-end, or even semantic feature-based methods. 
Notably on in-the-wild data, the proposed methods show drastic improvement over popular methods like AASIST, RawNet2, and others. 
The ablation study eliminates any doubt that performance improvements stem solely from the detector backend, confirming that non-semantic features are inherently better suited for generalization.
Despite these advancements, the quest for robust and generalized spoofing detection countermeasures remains ongoing.
Future work will explore alternative backend architectures, such as graph neural networks, in conjunction with these non-semantic feature extraction frontends.
Additionally, we will extend our study to partial spoofs and further investigate the comparison between semantic and non-semantic features by fine-tuning the frontend feature extractors for spoofing detection.
\section{Acknowledgements}
This research has been partly funded by the Federal Ministry of Education and Research, Germany (BMBF 03RU2U151C, project news-polygraph) and partly by the Volkswagen Foundation.

\bibliographystyle{IEEEtran}
\bibliography{trill_fake_bib}

\begin{thebibliography}{10}
\providecommand{\url}[1]{#1}
\csname url@samestyle\endcsname
\providecommand{\newblock}{\relax}
\providecommand{\bibinfo}[2]{#2}
\providecommand{\BIBentrySTDinterwordspacing}{\spaceskip=0pt\relax}
\providecommand{\BIBentryALTinterwordstretchfactor}{4}
\providecommand{\BIBentryALTinterwordspacing}{\spaceskip=\fontdimen2\font plus
\BIBentryALTinterwordstretchfactor\fontdimen3\font minus \fontdimen4\font\relax}
\providecommand{\BIBforeignlanguage}[2]{{%
\expandafter\ifx\csname l@#1\endcsname\relax
\typeout{** WARNING: IEEEtran.bst: No hyphenation pattern has been}%
\typeout{** loaded for the language `#1'. Using the pattern for}%
\typeout{** the default language instead.}%
\else
\language=\csname l@#1\endcsname
\fi
#2}}
\providecommand{\BIBdecl}{\relax}
\BIBdecl

\bibitem{wang2020asvspoof}
X.~Wang, J.~Yamagishi, M.~Todisco, H.~Delgado, A.~Nautsch, N.~Evans, M.~Sahidullah, V.~Vestman, T.~Kinnunen, K.~A. Lee \emph{et~al.}, ``{ASVspoof} 2019: A large-scale public database of synthesized, converted and replayed speech,'' \emph{Computer Speech \& Language}, vol.~64, p. 101114, 2020.

\bibitem{muller2022does}
N.~M. M{\"u}ller, P.~Czempin, F.~Dieckmann, A.~Froghyar, and K.~B{\"o}ttinger, ``Does audio deepfake detection generalize?'' \emph{arXiv preprint arXiv:2203.16263}, 2022.

\bibitem{chen2017resnet}
Z.~Chen, Z.~Xie, W.~Zhang, and X.~Xu, ``Resnet and model fusion for automatic spoofing detection.'' in \emph{Interspeech}, 2017, pp. 102--106.

\bibitem{li2021replay}
X.~Li, N.~Li, C.~Weng, X.~Liu, D.~Su, D.~Yu, and H.~Meng, ``Replay and synthetic speech detection with res2net architecture,'' in \emph{ICASSP 2021-2021 IEEE international conference on acoustics, speech and signal processing (ICASSP)}.\hskip 1em plus 0.5em minus 0.4em\relax IEEE, 2021, pp. 6354--6358.

\bibitem{zhang2021fake}
Z.~Zhang, X.~Yi, and X.~Zhao, ``Fake speech detection using residual network with transformer encoder,'' in \emph{Proceedings of the 2021 ACM workshop on information hiding and multimedia security}, 2021, pp. 13--22.

\bibitem{yi2023audio}
J.~Yi, C.~Wang, J.~Tao, X.~Zhang, C.~Y. Zhang, and Y.~Zhao, ``Audio deepfake detection: A survey,'' \emph{arXiv preprint arXiv:2308.14970}, 2023.

\bibitem{wang2015relative}
L.~Wang, Y.~Yoshida, Y.~Kawakami, and S.~Nakagawa, ``Relative phase information for detecting human speech and spoofed speech.'' in \emph{INTERSPEECH}, 2015, pp. 2092--2096.

\bibitem{chen2021ur}
X.~Chen, Y.~Zhang, G.~Zhu, and Z.~Duan, ``Ur channel-robust synthetic speech detection system for asvspoof 2021,'' \emph{arXiv preprint arXiv:2107.12018}, 2021.

\bibitem{sahidullah15_interspeech}
M.~Sahidullah, T.~Kinnunen, and C.~Hanilçi, ``{A comparison of features for synthetic speech detection},'' in \emph{Proc. Interspeech 2015}, 2015, pp. 2087--2091.

\bibitem{hua2021towards}
G.~Hua, A.~B.~J. Teoh, and H.~Zhang, ``Towards end-to-end synthetic speech detection,'' \emph{IEEE Signal Processing Letters}, vol.~28, pp. 1265--1269, 2021.

\bibitem{mo2022multi}
Y.~Mo and S.~Wang, ``Multi-task learning improves synthetic speech detection,'' in \emph{ICASSP 2022-2022 IEEE International Conference on Acoustics, Speech and Signal Processing (ICASSP)}.\hskip 1em plus 0.5em minus 0.4em\relax IEEE, 2022, pp. 6392--6396.

\bibitem{tak2021end}
H.~Tak, J.-w. Jung, J.~Patino, M.~Kamble, M.~Todisco, and N.~Evans, ``End-to-end spectro-temporal graph attention networks for speaker verification anti-spoofing and speech deepfake detection,'' \emph{arXiv preprint arXiv:2107.12710}, 2021.

\bibitem{ravanelli2018speaker}
M.~Ravanelli and Y.~Bengio, ``Speaker recognition from raw waveform with sincnet,'' in \emph{2018 IEEE spoken language technology workshop (SLT)}.\hskip 1em plus 0.5em minus 0.4em\relax IEEE, 2018, pp. 1021--1028.

\bibitem{li2023advanced}
J.~Li, Y.~Long, Y.~Li, and D.~Xu, ``Advanced rawnet2 with attention-based channel masking for synthetic speech detection,'' in \emph{Proc. INTERSPEECH}, vol. 2023, 2023, pp. 2788--2792.

\bibitem{jung2022aasist}
J.-w. Jung, H.-S. Heo, H.~Tak, H.-j. Shim, J.~S. Chung, B.-J. Lee, H.-J. Yu, and N.~Evans, ``{AASIST}: Audio anti-spoofing using integrated spectro-temporal graph attention networks,'' in \emph{ICASSP 2022-2022 IEEE international conference on acoustics, speech and signal processing (ICASSP)}.\hskip 1em plus 0.5em minus 0.4em\relax IEEE, 2022, pp. 6367--6371.

\bibitem{yang2024robust}
Y.~Yang, H.~Qin, H.~Zhou, C.~Wang, T.~Guo, K.~Han, and Y.~Wang, ``A robust audio deepfake detection system via multi-view feature,'' in \emph{ICASSP 2024-2024 IEEE International Conference on Acoustics, Speech and Signal Processing (ICASSP)}.\hskip 1em plus 0.5em minus 0.4em\relax IEEE, 2024, pp. 13\,131--13\,135.

\bibitem{babu2021xls}
A.~Babu, C.~Wang, A.~Tjandra, K.~Lakhotia, Q.~Xu, N.~Goyal, K.~Singh, P.~Von~Platen, Y.~Saraf, J.~Pino \emph{et~al.}, ``Xls-r: Self-supervised cross-lingual speech representation learning at scale,'' \emph{arXiv preprint arXiv:2111.09296}, 2021.

\bibitem{chen2022wavlm}
S.~Chen, C.~Wang, Z.~Chen, Y.~Wu, S.~Liu, Z.~Chen, J.~Li, N.~Kanda, T.~Yoshioka, X.~Xiao \emph{et~al.}, ``Wav{LM}: Large-scale self-supervised pre-training for full stack speech processing,'' \emph{IEEE Journal of Selected Topics in Signal Processing}, vol.~16, no.~6, pp. 1505--1518, 2022.

\bibitem{hsu2021hubert}
W.-N. Hsu, B.~Bolte, Y.-H.~H. Tsai, K.~Lakhotia, R.~Salakhutdinov, and A.~Mohamed, ``Hubert: Self-supervised speech representation learning by masked prediction of hidden units,'' \emph{IEEE/ACM transactions on audio, speech, and language processing}, vol.~29, pp. 3451--3460, 2021.

\bibitem{kulkarni2024exploring}
A.~Kulkarni, H.~M. Tran, A.~Kulkarni, S.~Dowerah, D.~Lolive, and M.~M. Doss, ``Exploring generalization to unseen audio data for spoofing: Insights from ssl models,'' in \emph{ASVSpoof workshop 2024}, 2024.

\bibitem{desplanques2020ecapa}
B.~Desplanques, J.~Thienpondt, and K.~Demuynck, ``Ecapa-tdnn: Emphasized channel attention, propagation and aggregation in tdnn based speaker verification,'' \emph{arXiv preprint arXiv:2005.07143}, 2020.

\bibitem{muller2024mlaad}
N.~M. M{\"u}ller, P.~Kawa, W.~H. Choong, E.~Casanova, E.~G{\"o}lge, T.~M{\"u}ller, P.~Syga, P.~Sperl, and K.~B{\"o}ttinger, ``{MLAAD}: The multi-language audio anti-spoofing dataset,'' \emph{arXiv preprint arXiv:2401.09512}, 2024.

\bibitem{shor2020towards}
J.~Shor, A.~Jansen, R.~Maor, O.~Lang, O.~Tuval, F.~d.~C. Quitry, M.~Tagliasacchi, I.~Shavitt, D.~Emanuel, and Y.~Haviv, ``Towards learning a universal non-semantic representation of speech,'' \emph{arXiv preprint arXiv:2002.12764}, 2020.

\bibitem{shor2022trillsson}
J.~Shor and S.~Venugopalan, ``Trillsson: Distilled universal paralinguistic speech representations,'' \emph{arXiv preprint arXiv:2203.00236}, 2022.

\bibitem{klambauer2017self}
G.~Klambauer, T.~Unterthiner, A.~Mayr, and S.~Hochreiter, ``Self-normalizing neural networks,'' \emph{Advances in neural information processing systems}, vol.~30, 2017.

\bibitem{yamagishi2019cstr}
J.~Yamagishi, C.~Veaux, K.~MacDonald \emph{et~al.}, ``{CSTR VCTK Corpus}: English multi-speaker corpus for {CSTR} voice cloning toolkit (version 0.92),'' \emph{University of Edinburgh. The Centre for Speech Technology Research (CSTR)}, 2019.

\bibitem{liu2023asvspoof}
X.~Liu, X.~Wang, M.~Sahidullah, J.~Patino, H.~Delgado, T.~Kinnunen, M.~Todisco, J.~Yamagishi, N.~Evans, A.~Nautsch \emph{et~al.}, ``A{SV}spoof 2021: Towards spoofed and deepfake speech detection in the wild,'' \emph{IEEE/ACM Transactions on Audio, Speech, and Language Processing}, 2023.

\bibitem{wang2021investigating}
X.~Wang and J.~Yamagishi, ``Investigating self-supervised front ends for speech spoofing countermeasures,'' \emph{arXiv preprint arXiv:2111.07725}, 2021.

\bibitem{liu2023leveraging}
X.~Liu, M.~Liu, L.~Wang, K.~A. Lee, H.~Zhang, and J.~Dang, ``Leveraging positional-related local-global dependency for synthetic speech detection,'' in \emph{ICASSP 2023-2023 IEEE International Conference on Acoustics, Speech and Signal Processing (ICASSP)}.\hskip 1em plus 0.5em minus 0.4em\relax IEEE, 2023, pp. 1--5.

\bibitem{yan2024df40}
Z.~Yan, T.~Yao, S.~Chen, Y.~Zhao, X.~Fu, J.~Zhu, D.~Luo, C.~Wang, S.~Ding, Y.~Wu \emph{et~al.}, ``Df40: Toward next-generation deepfake detection,'' \emph{arXiv preprint arXiv:2406.13495}, 2024.

\bibitem{zhang2024audio}
Q.~Zhang, S.~Wen, and T.~Hu, ``Audio deepfake detection with self-supervised xls-r and sls classifier,'' in \emph{Proceedings of the 32nd ACM International Conference on Multimedia}, 2024, pp. 6765--6773.

\bibitem{chen2022speechformer}
W.~Chen, X.~Xing, X.~Xu, J.~Pang, and L.~Du, ``Speechformer: A hierarchical efficient framework incorporating the characteristics of speech,'' \emph{arXiv preprint arXiv:2203.03812}, 2022.

\bibitem{campbell1989syllable}
W.~N. Campbell, ``Syllable-level duration determination.'' in \emph{EUROSPEECH}, 1989, pp. 2698--2701.

\end{thebibliography}

\end{document}